\documentclass{article}
\usepackage{graphicx}
\usepackage{amsmath, amsthm, latexsym}
\usepackage{amssymb}
\begin{document}

\vspace*{3cm} \thispagestyle{empty}
\noindent \textbf{{\Large LETTER}}\\
\vspace{5mm}

\noindent \textbf{\Large A Simple Criterion for Nonrotating Reference Frames}\\

\noindent \textbf{\normalsize Peter Collas}\footnote{Department of Physics and Astronomy, California
State University, Northridge, Northridge, CA 91330-8268. Email: peter.collas@csun.edu.}
\textbf{\normalsize and David Klein}\footnote{Department of Mathematics, California State University,
Northridge, Northridge, CA 91330-8313. Email: david.klein@csun.edu.}\\

\vspace{4mm} \parbox{11cm}{\noindent{\small We prove a theorem that gives an easily 
verifiable necessary and sufficient condition for a reference frame with fixed spacelike coordinates 
to be nonrotating in the sense of Walker. Applications are discussed.}\vspace{5mm}}\\\vspace{6cm}
\pagebreak

\setlength{\textwidth}{27pc}
\setlength{\textheight}{43pc}
\noindent \textbf{{\normalsize 1. INTRODUCTION}}\\

\noindent Walker [1] defined the reference frame of a nonrotating observer as a frame in which the
acceleration of a free test particle in the neighborhood of the observer is independent of
its velocity. This definition is consistent with the corresponding definition in Newtonian mechanics, 
and is useful in general relativity. Nonrotating frames are essential tools in descriptions of 
relativistic precession [2] and frame dragging [3]. 

Walker showed that an orthonormal tetrad, at each point on a timelike path $\textbf{x}(t)$, formed by
the unit tangent vector to $\textbf{x}$ and the unit vectors in the directions of the space axes, is
nonrotating when each of the vectors in the tetrad satisfies Eq. (1) below. 

In this paper we prove a theorem that gives a simple necessary and sufficient condition for a 
coordinate frame to be nonrotating on a spacetime path $\textbf{x}(t)$ whose space coordinates are 
fixed and whose time coordinate varies.  The coordinate frame consists of the tangent vectors of the
coordinates in which the metric is given. Our theorem gives simple criteria for this coordinate frame
to be nonrotating at a given fixed point in space. The assumption that the space coordinates are fixed
along  the spacetime path is not as restrictive as it might appear. By
transforming to a new coordinate system, which is moving relative to the old one, a point with fixed
spatial coordinates in the new system corresponds to a path in space in the old system. Fermi 
coordinates along a geodesic provide such an example which we discuss in relation to our theorem 
in Section 3 below. We also illustrate how the 
theorem may be used by identifying a nonrotating frame for the (interior) van Stockum [4] metric
for an  infinite rotating dust cylinder. In Section 2 we state and prove the theorem. \\

\noindent \textbf{{\normalsize 2. THEOREM ON NONROTATING FRAMES }}\\

\indent The Fermi-Walker equations for a vector $\xi^{\alpha}$ are
\begin{equation}
\nabla_{\vec{u}}\;\xi^{\alpha}=-\Omega^{{\alpha}}_{\;\,\beta}\xi^{\beta}\,,
\end{equation}
 
\noindent where $\vec{u}$ is the four-velocity, $\Omega^{{\alpha}}_{\;\,\beta}=a^{\alpha}u_{\beta}- 
u^{\alpha} a_{\beta}$, and $a^{\alpha}$ is the four-acceleration. Throughout,
Greek indices take values from the set $\{0,1,2,3\}$, while Latin indices take values from the
set $\{1,2,3\}$.

Eq. (1) may be rewritten as 
\begin{equation}
\frac{d\xi^{\alpha}}{d\tau}= -\left(\Gamma^{\alpha}_{\;\,\beta\gamma}u^{\gamma}+\Omega^{\alpha}_{\;\,
\beta}\right)\xi^{\beta}\,,
\end{equation}

\noindent where $\tau$ is proper time. Let $x^{0}=t$, $x^{1}$, $x^{2}$, $x^{3}$ be 
coordinates for a chart $U$ of spacetime. 
We consider the trajectory in spacetime given by  
$\textbf{x}(t)=(t,x^{1}_{0},x^{2}_{0},x^{3}_{0})$, 
where $x^{1}_{0},x^{2}_{0},x^{3}_{0}$ are the coordinates of a fixed point in space and
$(t,x^{1}_{0},x^{2}_{0},x^{3}_{0})\in U$ for all $t$ in 
an open interval $I$ on the real line\\
\indent Assume the following conditions on 
the metric $g =g_{\alpha \beta} dx^{\alpha}dx^{\beta}$:\\

\indent \textbf{(a)} For the metric $g$ on $\textbf{x}(t)$ we have that
\[g(\textbf{x}(t)) = g_{00}(\textbf{x}(t))(dx^{0})^2 
+ g_{11}(\textbf{x}(t))(dx^{1})^2
+ g_{22}(\textbf{x}(t))(dx^{2})^2 + g_{33}(\textbf{x}(t))(dx^{3})^2,\] 
where $g_{00}(\textbf{x}(t))<0$
and the other three coefficients are positive, for $t\in I$.  In other words, 
we assume that $g_{\alpha \beta}(\textbf{x}(t))=0$ when
$\alpha \neq \beta$ (though  $g_{\alpha \beta}$ may be nonzero on $U-\{\textbf{x}(t)\}$)
and that the coordinate $x_{0}=t$ is the time coordinate on $\textbf{x}(t)$.\\
\indent \textbf{(b)} $g_{\alpha \beta, 0}=0$ on the path 
$\textbf{x}(t)$ for all $\alpha$ and $\beta$.  In other words, the metric is stationary.\\
\indent In more concise terms these conditions require $g$ to be diagonal
and stationary on the spacetime path $\textbf{x}(t)$. With these assumptions, the four-velocity 
of the observer $\textbf{x}(t)$ may now be expressed as 
$\vec{u}=(dt/d\tau,dx^{1}_{0}/d\tau,dx^{2}_{0}/d\tau,dx^{3}_{0}/d\tau)=
(1/\sqrt{-g_{00}(\textbf{x}(t))}, 0, 0,0)$.\\

\noindent \textbf{Theorem}:  \textit{Assume that the metric $g$ satisfies conditions (a) and (b) 
on} $\textbf{x}(t)$. \textit{Then the orthonormal coordinate frame given by}
$\vec{u}$ \textit{together with the three spacelike vectors} $1/\sqrt{g_{ii}(\textbf{x}(t))}\;
\partial/ \partial x^{i}$ \textit{is non rotating along the path} $\textbf{x}(t)$ \textit{(in 
the sense that each vector field satisfies the Fermi-Walker equations) if and only if}
$g_{i0,j}=g_{j0,i}$ \textit{on} $\textbf{x}(t)$ \textit{for all} $i,j=1,2,3$.\\

\noindent \textbf{Remark} The Theorem may be reformulated to say that the coordinate tangent 
vectors form a nonrotating orthonormal frame if and only if the curl of the vector $(g_{10}, g_{20},
g_{30})$ vanishes, i.e., if and only if the one-form $g_{i0}dx^{i}$ is closed on $\textbf{x}(t)$.

The proof of the theorem depends on the following observations and a lemma.  If $g$ satisfies (a) 
and (b), then on $\textbf{x}(t)$
\begin{eqnarray}
\Gamma^{\alpha}_{\;\,\beta\gamma}&=&\frac{1}{2}g^{\alpha\mu}(g_{\mu\beta,\gamma}+
g_{\mu\gamma,\beta}-g_{\beta\gamma,\mu})\nonumber\\
&=&\frac{1}{2}g^{\alpha\alpha}(g_{\alpha\beta,\gamma}+g_{\alpha\gamma,\beta}-
g_{\beta\gamma,\alpha})\,,\;\;\;\mbox{(no sum)}\,.
\end{eqnarray}
The relations below follow immediately:
\begin{eqnarray}
\Gamma^{0}_{\;\,j0}&=&\frac{1}{2}g^{00}g_{00,j}\,,\\
\Gamma^{j}_{\;\,00}&=&-\frac{1}{2}g^{jj}g_{00,j}\,,\\
\Gamma^{i}_{\;\,i0}&=&0\,.
\end{eqnarray}\\
\noindent \textbf{Lemma}: \textit{If $g$ satisfies conditions (a) and (b) on} $\textbf{x}(t)$,
\textit{then}:

(1) \textit{For all} $\alpha$,
\begin{equation}
\frac{du^{\alpha}}{d\tau}=0\,.
\end{equation}

(2) \textit{The acceleration components are given by} 
\begin{equation}
a^{\alpha}=-\frac{\Gamma^{\alpha}_{\;\,00}}{g_{00}}\,.
\end{equation}

(3) \textit{On} $\textbf{x}(t)$ \textit {the tensor} $\Omega^{{\alpha}}_{\;\,\beta}=a^{\alpha}
u_{\beta}-u^{\alpha} a_{\beta}$, \textit{satisfies}:
\begin{eqnarray}
\Omega^{0}_{\;\,0}&=&0\,,\\
\Omega^{{\alpha}}_{\;\,\beta}&=&0\,,\;\;\mbox{\textit{unless}}\;\;\alpha\;\;\mbox{or}\;\;
\beta=0\,,\\\Omega^{0}_{\;\,j}&=&-g_{jj}u^{0}a^{j}\,.
\end{eqnarray}

\noindent\textbf{Proof of the Lemma}.  For part (1) observe that, 
\begin{equation}
\frac{du^{\alpha}}{d\tau}=
\frac{\partial u^{\alpha}}{\partial x^{\mu}}
\frac{\partial x^{\mu}}{\partial \tau} =
\frac{\partial u^{\alpha}}{\partial x^{0}}u^{0}=0,
\end{equation}
because $u^{0}=1/\sqrt{-g_{00}}$ does not depend on $x^{0}=t$
and the other components of $\vec{u}$ are zero.  Part (2) follows from

\begin{equation}
a^{\alpha}=\frac{du^{\alpha}}{d\tau}+\Gamma^{\alpha}_{\;\,\beta\gamma}
u^{\beta}u^{\gamma}=\Gamma^{\alpha}_{\;\,00}u^{0}u^{0},
\end{equation}
where the second equation is a consequence of part (1). For part (3), observe that because $g$ is
diagonal on  $\textbf{x}(t)$, $\Omega^{0}_{\;\,0}
=a^{0}u_{0}-u^{0}a_{0}=g_{00}(a^{0}u^{0}-u^{0}a^{0})=0$. The tensor
$\Omega^{{\alpha}}_{\;\,\beta}=a^{\alpha}u_{\beta}-  u^{\alpha} a_{\beta}$ is zero unless 
$\alpha$ or $\beta=0$  because $u^{j}$ and $u_{j}$ vanish. Finally,
$\Omega^{0}_{\;\,j}= a^{0}u_{j}-u^{0}a_{j}=-u^{0}a_{j}=-g_{jj}u^{0}a^{j}$.\\

\noindent\textbf{Proof of the Theorem}.  It is easy to check that any four-velocity automatically
satisfies the Fermi-Walker equations. In component form, the spacelike vectors of the tetrad are 
$(0,1/\sqrt{g_{11}},0,0), (0,0,1/\sqrt{g_{22}},0),(0,0,0,1/\sqrt{g_{33}})$. Choose any one of 
these vectors and denote it by $\vec{\xi} = (\xi^{0},\xi^{1},\xi^{2},\xi^{3})$. 
We first show that if $g$ satisfies conditions (a) and (b), then
\begin{equation}
\frac{d\xi^{0}}{d\tau}= -\left(\Gamma^{0}_{\;\,\beta\gamma}u^{\gamma}+
\Omega^{0}_{\;\,\beta}\right)\xi^{\beta}\,.
\end{equation}
\noindent Each step of the calculation below follows from Eqs. (4) - (6), the Lemma,
or from fact that $g^{jj}=1/g_{jj}$ for a diagonal matrix $g$.
\begin{eqnarray}
-\left(\Gamma^{0}_{\;\,\beta\gamma}u^{\gamma}+\Omega^{0}_{\;\,\beta}\right)\xi^{\beta}
&=&-\Gamma^{0}_{\;\,\beta0}u^{0}\xi^{\beta}-\Omega^{0}_{\;\,j}\xi^{j}\nonumber\\
&=&-\Gamma^{0}_{\;\,j0}u^{0}\xi^{j}+g_{jj}u^{0}a^{j}\xi^{j}\nonumber\\
&=& \left(-\Gamma^{0}_{\;\,j0}+g_{jj}a^{j}\right)u^{0}\xi^{j}\nonumber\\
&=& \left(-\frac{1}{2} g^{00}g_{00,j}-g_{jj}\frac{\Gamma^{j}_{\;\,00}}
{g_{00}}\right)u^{0}\xi^{j}\\
&=& \left(-\frac{1}{2} g^{00}g_{00,j}+g_{jj}\frac{\frac{1}{2} g^{jj}g_{00,j}}
{g_{00}}\right)u^{0}\xi^{j}\nonumber\\
&=&0=\frac{d\xi^{0}}{d\tau}\,.\nonumber
\end{eqnarray}
\noindent For the spatial components $(j=1,2,3)$,
\begin{equation}
d\xi^{j}/d\tau=0. 
\end{equation}

\noindent Eq. (16) follows from the same argument as in the proof of part 1 of the Lemma. 
On the other hand, using Eqs. (4) - (6), the Lemma, and  $\xi^{0}=0$, we find that,
\begin{eqnarray}
-\left(\Gamma^{j}_{\;\,\beta\gamma}u^{\gamma}+\Omega^{j}_{\;\,
\beta}\right)\xi^{\beta}
&=&-\Gamma^{j}_{\;\,\beta0}u^{0}\xi^{\beta}-\Omega^{j}_{\;\,
\beta}\xi^{\beta} \nonumber\\
&=&-\Gamma^{j}_{\;\,\beta 0}u^{0}\xi^{\beta}-\Omega^{j}_{\;\,
0}\xi^{0}\nonumber\\
&=&-\Gamma^{j}_{\;\,i 0}u^{0}\xi^{i}\\
&=&\frac{1}{2}g^{jj}(g_{ji,0}+g_{j0,i}-
g_{i0,j})u^{0}\xi^{i}\nonumber\\&=&\frac{1}{2}g^{jj}(g_{j0,i}-
g_{i0,j})u^{0}\xi^{i}.\nonumber\\
\nonumber
\end{eqnarray}

\noindent In view of Eqs. (16) and (17), the Fermi-Walker equations (2) are satisfied on
$\textbf{x}(t)$ for all $\alpha$ if and only if $g_{j0,i}-g_{i0,j}=0$ for all $i,j$. This 
concludes the proof of the theorem.\\

\noindent \textbf{{\normalsize 3. THE VAN STOCKUM METRIC}}\\

We illustrate the above theorem by applying it to the van Stockum metric [2]. The van Stockum 
solution represents a rotating dust cylinder of infinite extent  along the axis of symmetry ($z$-axis)
but of finite radius.  We consider only the interior solution which in coordinates rotating with the
dust particles takes its simplest form below:
\begin{equation}
ds^{2}=-Fdt^{2}+Ld\phi^{2}+2Mdtd\phi+Hdr^{2}+Hdz^{2}\;,
\end{equation}
where,
\begin{equation}
F=1\,,\;\;\;\;L=r^{2}(1-a^{2}r^{2})\,,\;\;\;\;M=ar^{2}\,,\;\;\;\;
H=e^{-a^{2}r^{2}}\,.
\end{equation}

\noindent In Eqs. (19), $0\leq r \leq R$ for a constant $R$ that determines the radius of the
cylinder, and $a$ is the angular velocity of the dust particles.

We now perform a rotation to noncomoving coordinates by the transformation
\begin{equation}
t=\bar{t}\;,\;\;\;\;\phi=\bar{\phi}-\Omega \bar{t}\;,\;\;\;\;r=\bar{r}\;,\;\;\;\;z=\bar{z}\;.
\end{equation}

\noindent In the barred coordinates, the metric
coefficients are:
\begin{equation}
\bar{F}=F+2\Omega
M-\Omega^{2}L\;,\;\;\;\bar{L}=L\;,\;\;\;\bar{M}=M-\Omega L\;,\;\;\;\bar{H}=H\;.
\end{equation} 
\noindent In order to eliminate the coordinate singularity at $\bar{r}=0$, we 
change to Cartesian coordinates by the transformation 
\begin{equation}
\bar{r}^{2} = x^{2}+y^{2},\;\;\;\tan\bar{\phi}=\frac{y}{x},\;\;\;
d\bar{r}=\frac{xdx+ydy}{\sqrt{x^{2}+y^{2}}},\;\;\;
d\bar{\phi}=\frac{xdy-ydx}{x^{2}+y^{2}}\,.
\end{equation}

\noindent Substituting and collecting terms we obtain,
\begin{eqnarray}
ds^{2}&=&-\bar{F}dt^{2}+\left(\frac{\bar{L}}{r^{4}}y^{2}+\frac{\bar{H}}{r^{2}}x^{2}\right)dx^{2}
+\left(\frac{\bar{L}}{r^{4}}x^{2}+\frac{\bar{H}}{r^{2}}y^{2}\right)dy^{2}\nonumber\\
&+& 2\left(\frac{\bar{H}}{r^{2}}-\frac{\bar{L}}{r^{4}}\right)xy\,dxdy\nonumber\\
&+&2\frac{\bar{M}}{r^{2}}x\,dtdy-2\frac{\bar{M}}{r^{2}}y\,dtdx+ \bar{H}dz^{2}\,,
\end{eqnarray}
and using Eqs. (21) we write 
\begin{eqnarray}
ds^{2}&=&-(F+2\Omega M-\Omega^{2}
L)dt^{2}+\left(\frac{L}{r^{4}}y^{2}+\frac{H}{r^{2}}x^{2}\right)dx^{2}\nonumber\\
&+&\left(\frac{L}{r^{4}}x^{2}+\frac{H}{r^{2}}y^{2}\right)dy^{2}+
2\left(\frac{H}{r^{2}}-\frac{L}{r^{4}}\right)xy\,dxdy\nonumber\\
&+&2\frac{(M-\Omega L)}{r^{2}}x\,dtdy-2\frac{(M-\Omega L)}{r^{2}}y\,dtdx+ Hdz^{2}\,.
\end{eqnarray}

\noindent Eqs. (19) may be substituted into this last expression, and the limits
as $r\rightarrow 0$ of each of the coefficients exists and they give the metric
along the axis of symmetry.  It is easily established that this metric is diagonal
on the axis of symmetry and in fact it equals the Minkowski metric on that axis, so conditions  (a)
and (b) are satisfied for all $\Omega$.  
An easy calculation shows that on the $z$-axis, 
\begin{equation}
g_{tx,y}=-g_{ty,x}=\Omega-a\,.
\end{equation}
By our theorem, the frame on the axis given by
$\partial/\partial t$, $\partial/\partial x$, $\partial/\partial y$,
$\partial/\partial z$ is nonrotating if and only if $\Omega=a$. This result was noted in [4]
from direct calculation using the Fermi-Walker equation, and a limiting argument.\\
\indent We note that the hypotheses of our theorem are satisfied by Fermi coordinates for an 
observer moving along a geodesic, as one would expect since the reference frame is inertial and
nonrotating by construction.  On a geodesic, the metric is Minkowskian in Fermi
coordinates, and the connection coefficients are zero along the path.  It follows immediately
that all first order partial derivatives of the metric coefficients vanish [5], and hence our
condition $g_{i0,j}=g_{j0,i}$ is satisfied trivially.\\

\noindent \textbf{ACKNOWLEDGMENT}\\

\noindent The authors wish to thank Professor John Lawrence for helpful comments.\\

\noindent \textbf{{\normalsize REFERENCES}}

\begin{enumerate}
\def\labelenumi{[\theenumi]}
\item Walker, A. G. (1935). \textit{Proc. Edin. Math. Soc.} \textbf{4}, 170.
\item Hamilton, J. D. (1996). \textit {Am. J. Phys.} \textbf{64}, 1197.
\item Collas, P. and Klein, D. (2004). \textit{Gen. Rel. Grav. }\textbf{36}, 1197.
\item van Stockum, W. J. (1937). \textit{Proc. R. Soc. Edin.} \textbf{57}, 135.
\item Misner, C. W., Thorne, K. S., and Wheeler, J. A. (1973). \textit{Gravitation}, W. H.
Freeman, San Francisco, p. 331.
\end{enumerate}

\end{document}